\documentclass[12pt]{iopart}

\usepackage{graphicx}
\DeclareGraphicsExtensions{.png,.jpg,.eps}
\usepackage{xcolor}

\begin{document}

\title{Unconventional  Yu-Shiba-Rusinov states in hydrogenated graphene}

\author{J. L. Lado$^1$ and J. Fern\'andez-Rossier$^{1,2}$}
\address{  (1) International Iberian Nanotechnology Laboratory (INL),
Av. Mestre Jos\'e Veiga, 4715-330 Braga, Portugal
\\
(2)  Departamento de F\'isica Aplicada, Universidad de Alicante, 03690 Spain
}

\date{\today} 

\begin{abstract} 

Conventional in-gap Yu-Shiba-Rusinov states require two
ingredients: magnetic atoms and a superconducting host that, in the
normal phase, has a finite density of states at the Fermi energy.  Here we show that hydrogenated graphene 
can host Yu-Shiba-Rusinov states  without any of those two ingredients.  Atomic hydrogen chemisorbed in graphene is known to act as paramagnetic center with a weakly localized  magnetic moment. Our calculations for 
hydrogenated graphene in proximity to a superconductor show that individual adatoms   induce  in-gap Yu-Shiba-Rusinov states with an exotic spectrum 
whereas  chains of adatoms result in a gapless Yu-Shiba-Rusinov band.   Our predictions can be tested using  state of the art techniques, combining recent progress of atomic manipulation of atomic hydrogen on graphene together with  the well tested proximity effect in graphene.
\end{abstract}
\maketitle

\ioptwocol

\section{Introduction}

Magnetic moments have long  been
known\cite{suhl1959impurity,shiba1968classical}
to deplete the superconducting order, 
hindering the coexistence of ferromagnetism and superconductivity. At the atomic scale, 
a single magnetic atom can  locally modify the superconducting order parameter,  binding in-gap Yu-Shiba-Rusinov (YSR)
states\cite{shiba1968classical,shiba1969superconducting,yu1965bound,rusinov1969theory}.
Using scanning tunneling spectroscopy (STS), 
these in-gap states have been observed  in a variety of systems\cite{yazdani97,ji2008high,hatter2015magnetic, menard2015coherent, ruby2015tunneling}, 
all of them involving transition metal or rare earth local moments.    In presence of either spin-orbit coupling or non-collinear magnetic order,
chains  of YSR impurities\cite{heimes2014majorana}
have been predicted to  result
in topological superconductivity,  whose fingerprint would be the emergence of
 zero energy Majorana\cite{elliott2015colloquium} edge states. The
recent observation\cite{nadj2014observation} 
of zero energy   end states by means of STS  in atomic
chains of Fe atoms on the surface of superconducting
Pb has been interpreted along this
line and has triggered and enormous interest in the
engineering of YSR states\cite{heimes2014majorana,pientka2013topological,poyhonen2014majorana}.

In conventional superconductors, the parameter that controls the energetics
 of YSR states is $\rho J$
 where $\rho$ is the density of states (DOS) at the
Fermi surface in the normal phase and $J$ is  strength of the
Kondo exchange between the local moment with spin $S$  and the
conduction electrons.  When $\rho$ is a slowly varying
function of energy,  the binding energy of YSR states
for  a classical spin in a superconductor,
is given by\cite{shiba1968classical,balatsky2006impurity}
\begin{equation}
E_S = \Delta \frac{1 - (\pi \frac{S}{2} J\rho)^2}{1 + (\pi \frac{S}{2}J\rho)^2}
\label{eqshiba}
\end{equation}
where $\Delta$ is the superconducting energy gap.
Thus,  according  to this classical result,   a finite $\rho$ is needed in order to have in-gap YSR states. 
We now address the question of how could graphene change this state of affairs.

\begin{figure}[h]
 \centering
                \includegraphics[width=.5\textwidth]{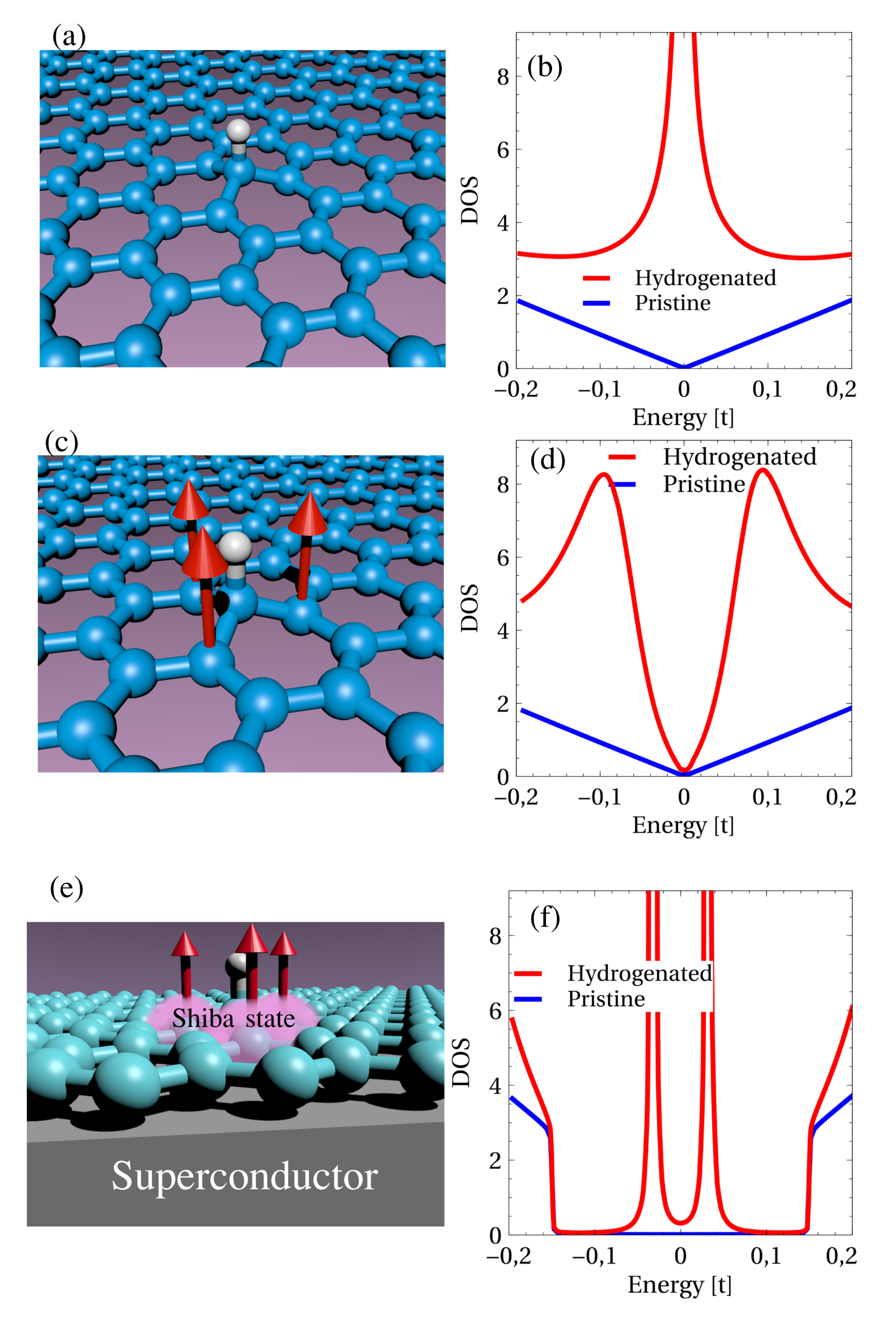}

\caption{ (a) Sketch of hydrogenated graphene (b) and DOS in the
carbons close to the hydrogen. Upon introduction
of interactions a local magnetic moment is developed (c),
changing dramatically the associated DOS (d).
When the system is placed on top of a superconductor
(e), YSR excitations
arise below the superconducting gap (f).
The parameters used are $\Delta=0.15t$ and $J=0.4$.
}
\label{doshy}
\end{figure}

Graphene can be made superconductor via proximity
effect
using several complementary strategies. 
On one hand superconductivity can be induced in lateral graphene/superconductor
heterostructures\cite{heersche2007bipolar,komatsu2012superconducting,calado2015ballistic} and on the other hand
taking advantage of its two dimensional character,  
graphene can be deposited on top of and beneath a
superconductor\cite{tonnoir2013induced,han2014collapse}.  The  recent reports of fabrication of vertical Van der Waals structures combining\cite{ugeda2015,cao2015quality,efetov2015specular} graphene with superconducting NbSe$_2$ also present a promising venue in this regard. Moreover,
a carbon layer of intercalated
graphite $C_6Ca$,\cite{yang2014superconducting,chapman2015superconductivity,rahnejat2011charge}
could be 
considered as superconducting
graphene,
although in the
former case the Fermi level is away from half filling.

Local magnetic moments can be induced in graphene without using transition metals
via chemisorption of  atomic hydrogen\cite{yazyev2007defect,Palacios08} as well as many other  covalent functionalizations\cite{nair12,santos2012universal}.
Within a one-body picture, 
the chemisorption of atomic hydrogen
in graphene creates a zero 
energy state,
which 
greatly enhances
the local DOS close to the Fermi energy. 
Electron-electron interactions  result\cite{yazyev2007defect,boukhvalov2008hydrogen,Palacios08,soriano2010hydrogenated} in the formation of a local moment associated to
chemisorbed hydrogen.
When two hydrogen atoms
are chemisorbed on the same sub lattice,  ferromagnetic couplings are 
expected\cite{Palacios08,soriano2010hydrogenated}.
These theoretical results are in line with  recent experiments\cite{HIvan} where both individual chemisorbed hydrogen, as well as dimers and trimers, have been probed using STS.  In these experiments
atomic manipulation of individual hydrogen atoms has been demonstrated, 
showing the potential for atomic scale engineering of magnetism in graphene.
In addition, this magnetism can be turned on and off when the density of carriers is changed \cite{HIvan}, 
in line with the  experimentally demonstrated  electrical control of paramagnetism in the case of
fluorinated graphene \cite{nair2013dual} and as expected from  
  theoretical calculations\cite{yndurain2014effect}.

\begin{figure}
 \centering
                \includegraphics[width=.5\textwidth]{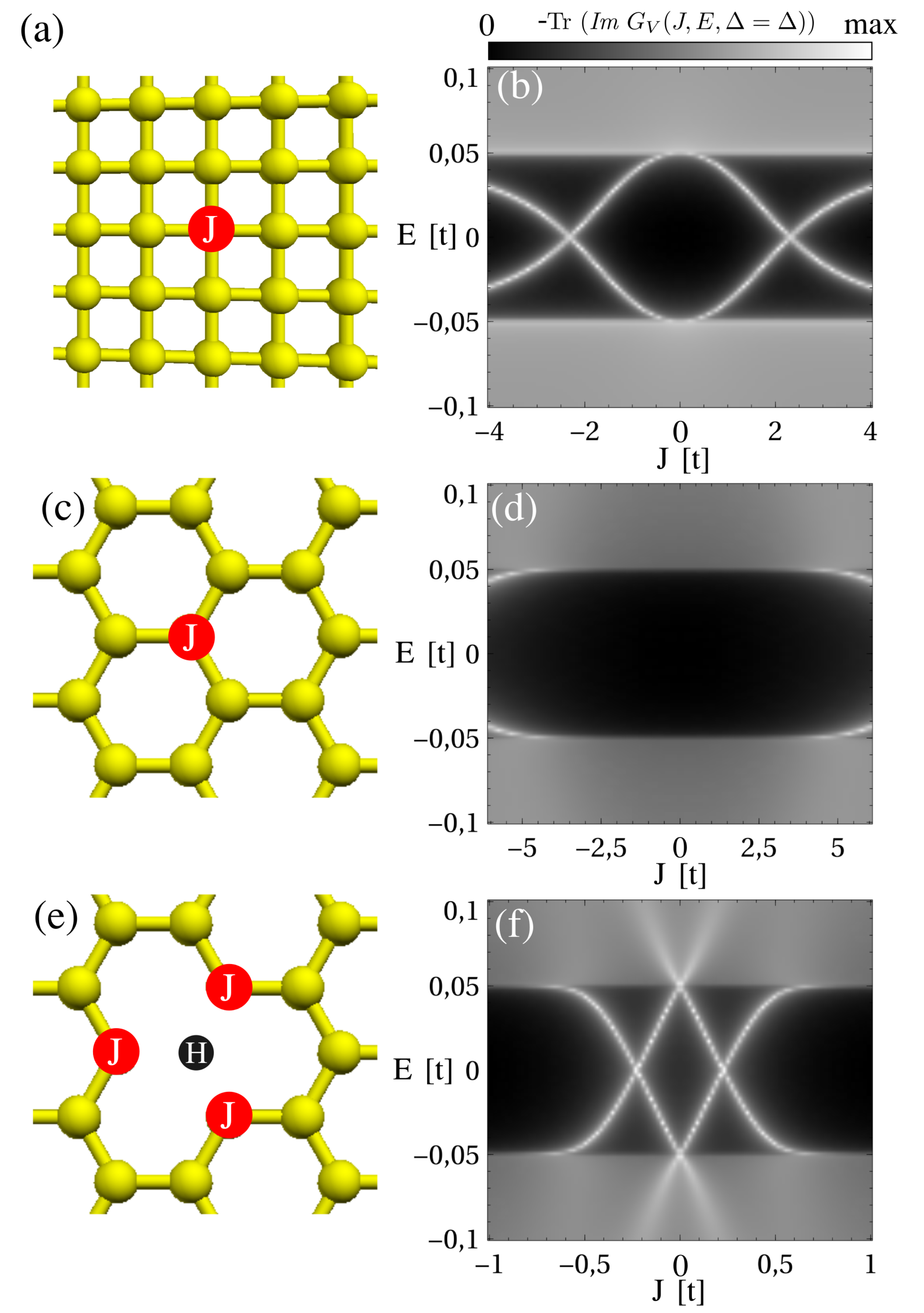}

\caption{
Single magnetic impurity
in an infinite square lattice (a), showing
in-gap states described by Eq.\ref{eqshiba}.
The same impurity in the 
honeycomb lattice (c) does not
show YSR states at weak coupling (d).
For the single hydrogenated graphene (e), 
a new and qualitatively different branch of in-gap
state arises (f).
}
\label{fig2}
\end{figure}


\section{Methods}

We now model hydrogenated graphene on top of a superconductor 
using a one orbital 
tight-binding model with pairing and
exchange fields.  Within the one-orbital model, the effect of hydrogenation
and other covalent functionalizations\cite{santos2012universal}
are equivalent
to the removal of a site in the lattice\cite{soriano2010hydrogenated}
without modifying the onsite energies of neighboring carbon atoms.
At the non-interacting level, this results in an in-gap $E=0$ state in the case of gapped graphene nanostructures, and a resonance in the case of 2D graphene
\cite{wehling2007local}.
In most instances,  interactions
have been included at a mean field level using
supercells\cite{yazyev2007defect}, that invariably result in spin-split solutions with a sublattice polarized  magnetization cloud in the neighborhood of the hydrogen atom and total spin $S=1/2$. 
Our Hamiltonian includes the spin-dependent potential in  the
 three closest carbon
atoms to the  one  underneath  the chemisorbed hydrogen.  This minimal model  mimics 
a self-consistently
calculated  exchange field that breaks time reversal symmetry that implies a  local magnetization
induced by the chemisorbed hydrogen. 
Finally, in order to account for the
proximity induced superconducting gap $\Delta$ , we include 
 a pairing term\cite{beenakker2006specular}   in the theory. Thus, the complete
  Bogoliubov-De Gennes (BdG)  Hamiltonian reads:
\begin{equation}
H = 
 H_{kin} + H_W+  H_{J} + H_{SC}
\label{full_ham}
\end{equation}

where $H_{kin}$  describes hopping, $H_W$ an onsite
potential term, $H_{J}$  is the exchange term
and $H_{SC}$ the superconducting  pairing. The hopping  term
is the standard nearest neighbor (NN) hopping: 

\begin{equation}
 H_{kin} = t \sum_s \sum_{(i,j) \in NN} c_{j\sigma}^\dagger c_{i\sigma}
\end{equation}

In the case of the hydrogenated system, the effect
of hydrogenation is captured by adding
an infinite  onsite energy $W$ to functionalized 
carbon site
\begin{equation}
H_{W} =  \lim_{W\rightarrow \infty} W \sum_s c^\dagger_{0,s}c_{0,s}
\end{equation}
The exchange term can be written down as:  
\begin{equation}
H_J =  \sum_{i } j(i) c_{i\sigma }^\dagger \frac{\sigma_z}{2} c_{i\sigma} 
\end{equation}

When we model the a conventional Shiba state in the square and honeycomb  lattice, we take $W=0$ and $j(i)$
takes a  non-zero value $J$ in just one site, marked  in red in figures 2(a,c).   In contrast, when we model 
 hydrogenated graphene we take  $W=\infty$ and $j(i)$ takes a non zero  value $J$       in the three first neighbors of the functionalized carbon site (see figure 2(e)).  We  treat   $J$ as a free parameter, to account for variations of the local magnetization coming from temperature or
doping\cite{nair2013dual,yndurain2014effect}.

Finally, the  superconducting proximity  effect is introduced as an
effective conventional
s-wave pairing term
\begin{equation}
H_{SC} =  \Delta \sum_{i} \left [ c_{i,\uparrow}c_{i,\downarrow} +
c^{\dagger}_{i,\downarrow} c^{\dagger}_{i,\uparrow} \right ]
\end{equation}

{Here $\Delta$ is taken as an input parameter in the calculation, and no attempt to compute it self-consistently is done}.
%
Since we are considering a single impurity in an infinite pristine system,
 we have to deal with a problem with infinite size and no
translational invariance.   We tackle the problem using  Green functions and a partition method, valid for any dimension. 
To do so,
we divide
the problem in a core region that contains
the impurity  site(s) 
and an outer region\cite{jacob2010orbital}.
This division is performed by creating a graphene supercell
as the new unit cell $C$, where the central supercell host
the hydrogenated site
and the sites with exchange coupling

\begin{equation}
h_V = h_{kin} + h_W + h_J + h_{SC} 
\end{equation}

with $ h_{kin}$, $h_W$, $h_J$ and $h_{SC}$ 
the projection of Eq. \ref{full_ham}
in the central defective supercell. The rest of the system is formed
of pristine supercells coupled to each other
and to the defective one. To calculate the full Green function
of the defective supercell, we write down the Dyson equation
of defective supercell coupled to the infinite graphene.
\begin{equation}
G_V (E)= (E - h_V - \Sigma (E))^{-1}
\label{dy_def}
\end{equation}
where $\Sigma(E)$ is the selfenergy induced over the defective supercell by the
rest of pristine system. The calculation of $\Sigma(E)$ is done 
noting that, for a pristine supercell,  an analogous Dyson equation can be written up:
%
\begin{equation}
G_0 (E)= (E - h_0 - \Sigma (E))^{-1}
\label{dy_pris}
\end{equation}
However, in the case of pristine graphene the Green function $G_0$
of the supercell
$C$ can also be calculated by summing up the Green functions
of the Bloch Hamiltonian $H_k$ as
\begin{equation}
G_0(E)= \frac{1}{(2\pi)^2}\int (E-H_k)^{-1} d^2 k
\label{bl_pris}
\end{equation}
with $H_k$ the usual Bloch Hamiltonian
\begin{equation}
H_k = h_0 + \sum_{\vec r} t_{\vec r} e^{i \vec k \cdot \vec r}
\end{equation}
where $h_0$ is the pristine intracell hopping matrix,
$t_{\vec r}$ are the intercell hopping matrices and $\vec r$ are 
real space vector connecting supercells. 
Coming back to Eq. \ref{dy_pris},
the self energy that pristine graphene induces on
a central supercell can be  calculated combining Eq. \ref{dy_pris}
and \ref{bl_pris} as
\begin{equation}
\Sigma (E) =
E - h_0 - 
\left ( \frac{1}{(2\pi)^2}\int (E-H_k)^{-1} d^2 k
\right )^{-1}
\label{self}
\end{equation}

Finally, with the previous self energy the full Green function
of the defective supercell $G_V$ can be calculated with Eq. \ref{dy_def}.
From the Green function, the
spectral function can easily be obtained as $\rho = -\frac{1}{\pi}\mathrm{Im} G_{V}(E)$
 introducing 
a small but finite imaginary part
 in the energy
$E\rightarrow E + i\delta$.
We stress that this method is specially well suited to capture
single impurities in infinite systems, not relying on periodic boundary
conditions and avoiding undesired interference effects between
different impurities.

\begin{figure}[t]
 \centering
                \includegraphics[width=.5\textwidth]{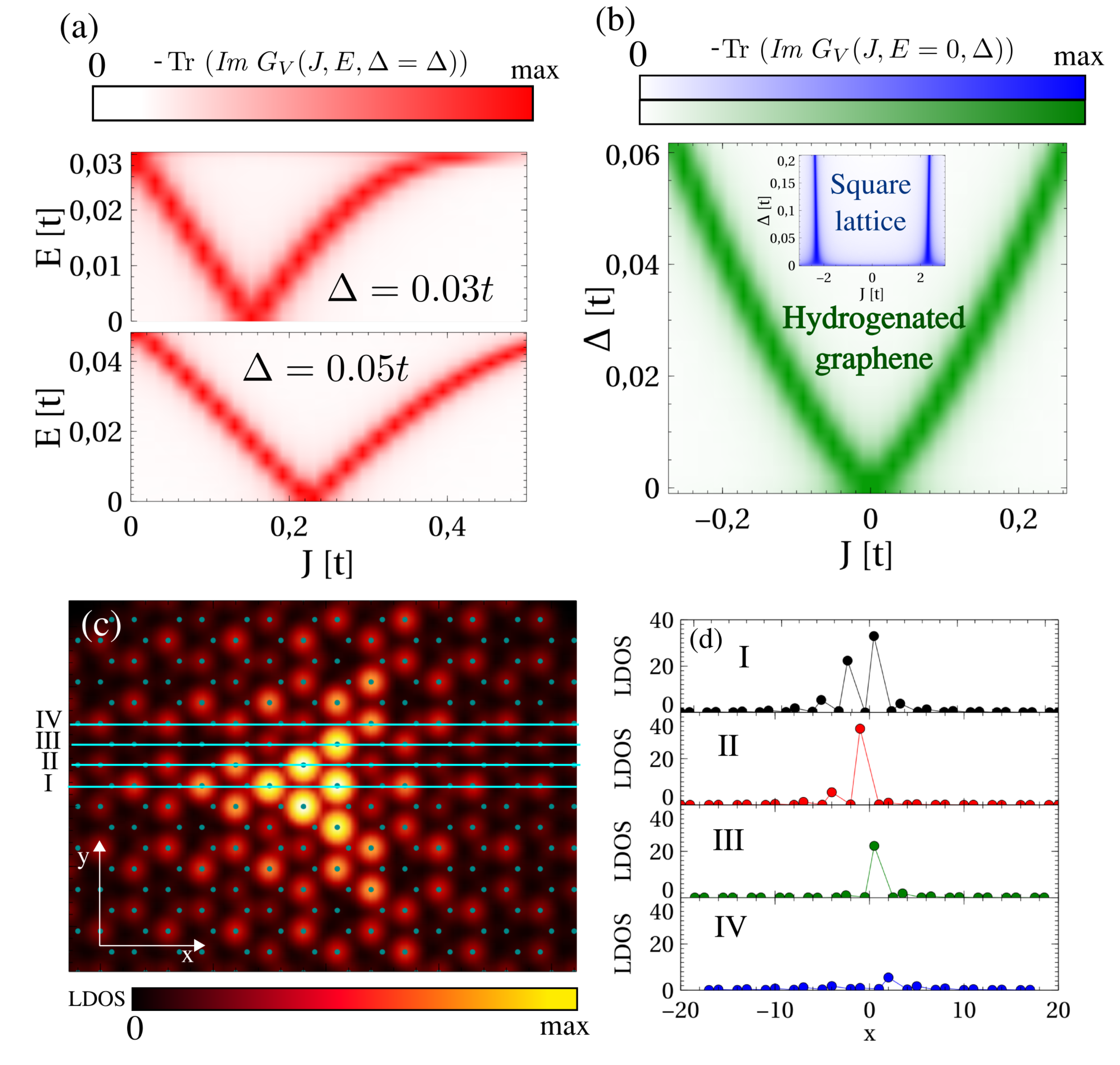}

\caption{ (a) Low energy spectral function of the YSR states
for single hydrogenated graphene, for different
paring couplings, showing an anomalous displacement of the pairing
switching point. (b) $E = 0$ spectral function as a function
of superconducting paring and exchange coupling for hydrogenated graphene
(green), showing a non linear
dependence of the switching point.
The inset (blue) shows the result for the usual parabolic band
which yields a $\Delta$ independent transition point. (c)
Spatially resolved DOS at $E=0$ for the parity switching point
$J=0.22t$, $\Delta=0.05t$. Panel (d) shows the amplitude
of the DOS along different lines, marked in panel (c), 
showing a strong localization of the YSR state close to
the hydrogenated site. In panel (c), the hydrogen adatom is deposited
in the center of the triangular pattern.

}
\label{pairing}
\end{figure}

\section{Results}

It is instructive to analyze the density of states
of the single hydrogenated graphene in three stages.
First,  with $J=\Delta=0$, we see in figure \ref{doshy}(b) how the density
of states diverges for $E=0$, in
line with analytic results \cite{peres2006electronic}.
Second, when the effect of the mean field
exchange is added ($J\neq 0$ in our
Hamiltonian), the  $E=0$ peak spin splits,  and results in a vanishing DOS at $E=0$  (see Fig.  \ref{doshy}(d)).
The resulting DOS calculated using
the embedding method 
shows a phenomenology analogous to a toy model, a 
zero energy level,  spin splitted by an exchange $J$ and coupled to a bath 
with the graphene density of states $\rho_0=\lambda |E|$. 
In this toy model, the Dyson equation gives the  spectral function 
$
\rho_{\pm} (E,J) = \frac{\lambda|E|+0^+}{(J\pm E)^2
+ (\lambda|E|+0^+)^2}
$.
The dramatic difference between the results with $J=0$ (Fig. 1b)
and $J\neq0$ (Fig. 1d) are analogous to the pathological
behavior of the function $\rho(E,J)$.
In particular, $\rho$ can not be Taylor
expanded for $E=0$, because the small $J$ and $E$ limits  can not be exchanged.
%
This  prevents the  use  of $\rho J$ as a well defined function and 
invalidates the use of   Eq.\ref{eqshiba} to model   YSR states in hydrogenated graphene.

Finally, in the third stage,
we study the effect of the  superconducting proximity effect on graphene. 
A proximity gap opens in the  DOS
of pristine graphene (blue line in Fig. 1(f)). In contrast, the
calculated DOS close to a chemisorbed hydrogen atom
shows an intra-gap   YSR state (red line in Fig. 1(f))

\subsection{YSR states for individual magnetic centers}

The in-gap excitation energy
is governed by the strength of the exchange coupling. Using our methodology for a magnetic impurity embedded in  square lattice (Fig. 2a) ,  with conventional 
  parabolic bands, the evolution of the in-gap YSR state as a function of $J$  is shown in Fig. 2b, with $\Delta=0.05t$ 
 taking chemical potential   $E_F=-2t$. Our results 
follow Eq. \ref{eqshiba}. In particular, in the low $J$ limit
the in-gap energies $E_S$ follow a quadratic law  $ \Delta-E_S \propto  J^2$. 

In contrast, a single magnetic impurity (Fig. 2c)  in the honeycomb lattice at half filling yields
no in-gap state unless the exchange interaction $J$ takes unrealistically large values $J>>t$ (Fig. 2d), in line with a previous result\cite{wehling2008local}.
This can be understood within the standard model as a straightforward consequence of 
 the vanishing DOS at $E=0$.  The situation is radically different when we consider the model for hydrogenated graphene (Fig. 2e).
In this case, in-gap YSR states appear at weak coupling that, in contrast with conventional YSR states, 
follow a linear evolution with $J$ at low coupling, and
therefore are not described by Eq. \ref{eqshiba}. This is the main result of this paper.

On top of their linear dependence on $J$, the hydrogenated graphene YSR states have another unconventional property.
Let us define  $J_c$ as the  point that satisfies  $E_S(J_c)=0$, which marks  a parity switching of the ground state
between a singlet state for $J<J_c$ to a doublet 
state for $J>J_c$.\cite{balatsky2006impurity}
For the conventional case, equation (1) shows how 
 $J_c$ 
is independent
from the superconducting pairing $\Delta$,
depending solely on
the density of states at Fermi energy {$\pi \frac{S}{2} \rho J_c =1$}. In comparison,
for hydrogenated graphene the parity switching point
is $\Delta$ dependent, as can be observed in Fig.\ref{pairing}(a).
In Fig. \ref{pairing}(b) we plot a contour map of the
BdG spectral function evaluated at $E=0$ as a function of $J$ and $\Delta$, showing a clear linear dependence of
$J_c$ on $\Delta$. 
In contrast, in the case of a square lattice, the same procedure yields
a $J_c$ that is  independent of $\Delta$ (inset of Fig.\ref{pairing}(b)). 

Thus, from an experimental
point of view, the parity switching point
could be observed by controlling either the superconducting gap $\Delta$, that depends strongly on temperature, as well as 
 tuning the hydrogen magnetic moment by controlling the doping level of graphene with a gate\cite{nair2013dual}.
Given that $E_S\simeq \Delta -  J/6$,  the critical $J_c$ is in the range of the  $6\Delta$,  ie, in the range of  10 meV\cite{mcmillan1968tunneling,chrestin1997evidence}.
Another prediction for experiments is shown in  figures 3(c) and 3(d), where we  show the  spectral function of the YSR states, as it would be measured with an STM.
This hydrogenated-graphene YSR wave function inherits
both the extension and the $C_3$ symmetry of the impurity resonance  of the normal phase\cite{yazyev2007defect}
 (Fig. 3(c)).  In particular, the YSR state peaks on the first neighbors of the
hydrogen atom (Fig. 3(d)).

\subsection{YSR states for superlattices}

\begin{figure}[t]
 \centering
                \includegraphics[width=.5\textwidth]{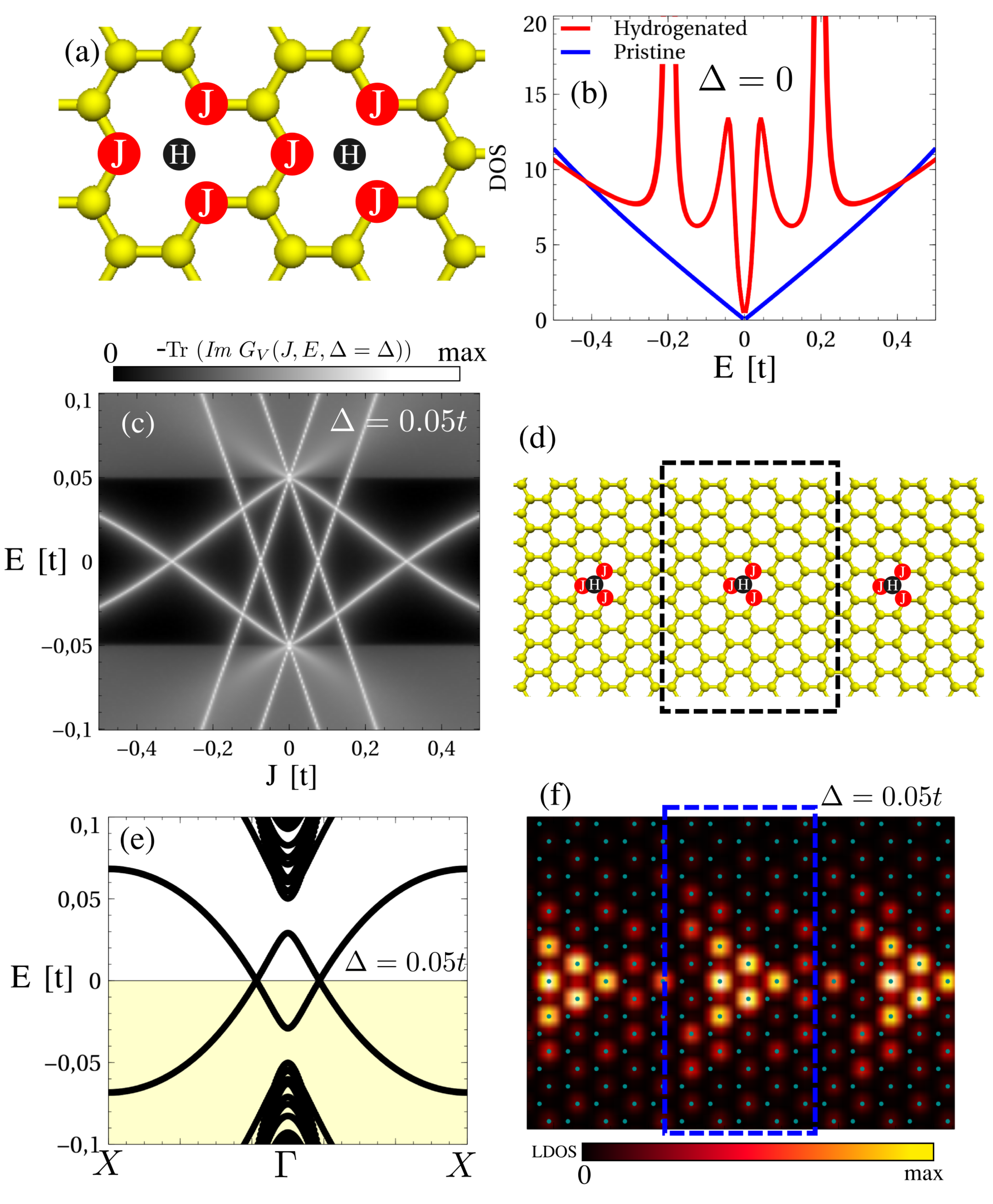}

\caption{ 
(a) Central cell with two hydrogen atoms, and (b) density of states
before the superconducting proximity is considered. (c) Evolution
of the YSR spectrum for  case (a), resembling the same linear evolution
as Fig. 2f. (d) Sketch of the unit cell for a periodic
array of hydrogenated sites. Panel (e) shows the BdG
band structure of the periodic array, showing a gapless YSR
band. Panel (f) shows the spatial
resolved DOS for the periodic array (d) at $E=0$. 
}
\label{many}
\end{figure}

We now consider YSR state superlattices formed by several hydrogen atoms
chemisorbed in the same sublattice. This secures a ferromagnetic coupling between them\cite{Palacios08,soriano2010hydrogenated}.  
We  first consider the case of a dimer (Fig. 4a) which  is expected\cite{Palacios08} to have $S=1$.  The resulting density of states with $\Delta=0$ and $J\neq0$
shows two peaks (Fig. 4b), rather than only one (Fig. 1d) for the single hydrogen case.
When the superconducting pairing is switched on, the YSR spectrum also acquires an extra line, compared to the single hydrogen case. 
The evolution of the two YSR states is still linear with $J$, as in the single hydrogen case.  Thus, there are as many bound YSR states as hydrogen atoms.  We now extend this notion to the case of a one dimensional periodic  array of hydrogen atoms (Fig. 4d). For this
one dimensional YSR crystal we obtain a single
YSR band that corresponds to a
gapless 1D superconductor, reminiscent of the one recently found at the interface of a magnetically ordered graphene edge and a superconductor\cite{san2015majorana}. 
The map of density of states at $E=0$ for the one dimensional array is shown in (Fig.\ref{many}f), whose spatial profile resembles
the single hydrogen YSR state.

So far we assumed that resonant magnetism behaves as a classical
magnetic moment. 
It must be noted  that quantum fluctuations scale as $\frac{1}{S}$ and are thus expected to have an important role
 both in  conventional
YSR states\cite{lim2015shiba,matsuura1977effects,nagaoka1971bound,soda1967sd,satori1992numerical,bauer2007spectral,kirvsanskas2015yu} as well as in the case of  hydrogenated graphene\cite{haase2011magnetic,sofo2012magnetic}.
Such fluctuations are not captured within the
present theoretical framework. However, our approach becomes more accurate for the larger structures considered in figure 4, that have a  larger spin, and thereby smaller quantum fluctuations.

\section{Conclusions}

We have shown that a single chemisorbed
hydrogen in superconducting graphene
creates a YSR bound state, in spite of the vanishing density of states
of pristine graphene. These YSR states have properties
very different from conventional YSR states, 
such as a  linear dependence of the binding energy  $E_S$ with exchange, and a 
parity switching point $J_c$ that depends on the
superconducting pairing energy, and  can thereby be modulated with
temperature. Motivated by recent experiments that demonstrate
the atomic manipulation of individual hydrogen atoms
on graphene\cite{HIvan}, we have also explored the
properties of YSR super-structures.
Furthermore, these results also apply for a much wider class
of covalent functionalizations in graphene\cite{santos2012universal}.
Combined with the electric control of magnetism, 
this  class of systems offers a unique platform to engineer
exotic superconducting states at the nanoscale.



\section{Acknowledgments}

 JFR acknowledges  financial supported by MEC-Spain (FIS2013-47328-C2-2-P) 
  and Generalitat Valenciana (ACOMP/2010/070), Prometeo. This work has
been financially supported in part by FEDER funds.  We acknowledge
financial support by Marie-Curie-ITN 607904-SPINOGRAPH. J. L. Lado thanks
the hospitality of the Departamento de Fisica Aplicada
at the Universidad de Alicante. We thank Ivan Brihuega for sharing experimental data. 
We thank D. Jacob for  bringing to our attention the embedding
method\cite{jacob2010orbital}.
We thank N. Garcia for insightful comments and proof reading of the manuscript.
 We thank  SPICE for support in the organization
of the workshop on "Magnetic adatoms as building blocks for quantum magnetism" that has inspired this work. 

\section*{References}
\bibliographystyle{unsrt}
\bibliography{biblio}{}

\begin{thebibliography}{10}

\bibitem{suhl1959impurity}
H~Suhl and BT~Matthias.
\newblock Impurity scattering in superconductors.
\newblock {\em Physical Review}, 114(4):977, 1959.

\bibitem{shiba1968classical}
Hiroyuki Shiba.
\newblock Classical spins in superconductors.
\newblock {\em Progress of theoretical Physics}, 40(3):435--451, 1968.

\bibitem{shiba1969superconducting}
Hiroyuki Shiba and Toshio Soda.
\newblock Superconducting tunneling through the barrier with paramagnetic
  impurities.
\newblock {\em Progress of Theoretical Physics}, 41(1):25--44, 1969.

\bibitem{yu1965bound}
L~Yu.
\newblock Bound state in superconductors with paramagnetic impurities.
\newblock {\em Acta Phys. Sin}, 21:75--91, 1965.

\bibitem{rusinov1969theory}
AI~Rusinov.
\newblock Theory of gapless superconductivity in alloys containing paramagnetic
  impurities.
\newblock {\em Sov. Phys. JETP}, 29(6):1101--1106, 1969.

\bibitem{yazdani97}
Ali Yazdani, B.~A. Jones, C.~P. Lutz, M.~F. Crommie, and D.~M. Eigler.
\newblock Probing the local effects of magnetic impurities on
  superconductivity.
\newblock 275(5307):1767--1770, 1997.

\bibitem{ji2008high}
Shuai-Hua Ji, Tong Zhang, Ying-Shuang Fu, Xi~Chen, Xu-Cun Ma, Jia Li, Wen-Hui
  Duan, Jin-Feng Jia, and Qi-Kun Xue.
\newblock High-resolution scanning tunneling spectroscopy of magnetic impurity
  induced bound states in the superconducting gap of pb thin films.
\newblock {\em Physical review letters}, 100(22):226801, 2008.

\bibitem{hatter2015magnetic}
Nino Hatter, Benjamin~W Heinrich, Michael Ruby, Jose~I Pascual, and Katharina~J
  Franke.
\newblock Magnetic anisotropy in shiba bound states across a quantum phase
  transition.
\newblock {\em Nature communications}, 6, 2015.

\bibitem{menard2015coherent}
Gerbold~C M{\'e}nard, S{\'e}bastien Guissart, Christophe Brun, St{\'e}phane
  Pons, Vasily~S Stolyarov, Fran{\c{c}}ois Debontridder, Matthieu~V Leclerc,
  Etienne Janod, Laurent Cario, Dimitri Roditchev, et~al.
\newblock Coherent long-range magnetic bound states in a superconductor.
\newblock {\em Nature Physics}, 11(12):1013--1016, 2015.

\bibitem{ruby2015tunneling}
Michael Ruby, Falko Pientka, Yang Peng, Felix von Oppen, Benjamin~W Heinrich,
  and Katharina~J Franke.
\newblock Tunneling processes into localized subgap states in superconductors.
\newblock {\em Physical review letters}, 115(8):087001, 2015.

\bibitem{heimes2014majorana}
Andreas Heimes, Panagiotis Kotetes, and Gerd Sch{\"o}n.
\newblock Majorana fermions from shiba states in an antiferromagnetic chain on
  top of a superconductor.
\newblock {\em Physical Review B}, 90(6):060507, 2014.

\bibitem{elliott2015colloquium}
Steven~R Elliott and Marcel Franz.
\newblock Colloquium: Majorana fermions in nuclear, particle, and solid-state
  physics.
\newblock {\em Reviews of Modern Physics}, 87(1):137, 2015.

\bibitem{nadj2014observation}
Stevan Nadj-Perge, Ilya~K Drozdov, Jian Li, Hua Chen, Sangjun Jeon, Jungpil
  Seo, Allan~H MacDonald, B~Andrei Bernevig, and Ali Yazdani.
\newblock Observation of majorana fermions in ferromagnetic atomic chains on a
  superconductor.
\newblock {\em Science}, 346(6209):602--607, 2014.

\bibitem{pientka2013topological}
Falko Pientka, Leonid~I Glazman, and Felix von Oppen.
\newblock Topological superconducting phase in helical shiba chains.
\newblock {\em Physical Review B}, 88(15):155420, 2013.

\bibitem{poyhonen2014majorana}
Kim P{\"o}yh{\"o}nen, Alex Weststr{\"o}m, Joel R{\"o}ntynen, and Teemu Ojanen.
\newblock Majorana states in helical shiba chains and ladders.
\newblock {\em Physical Review B}, 89(11):115109, 2014.

\bibitem{balatsky2006impurity}
AV~Balatsky, I~Vekhter, and Jian-Xin Zhu.
\newblock Impurity-induced states in conventional and unconventional
  superconductors.
\newblock {\em Reviews of Modern Physics}, 78(2):373, 2006.

\bibitem{heersche2007bipolar}
Hubert~B Heersche, Pablo Jarillo-Herrero, Jeroen~B Oostinga, Lieven~MK
  Vandersypen, and Alberto~F Morpurgo.
\newblock Bipolar supercurrent in graphene.
\newblock {\em Nature}, 446(7131):56--59, 2007.

\bibitem{komatsu2012superconducting}
Katsuyoshi Komatsu, Chuan Li, S~Autier-Laurent, H~Bouchiat, and S~Gu{\'e}ron.
\newblock Superconducting proximity effect in long
  superconductor/graphene/superconductor junctions: From specular andreev
  reflection at zero field to the quantum hall regime.
\newblock {\em Physical Review B}, 86(11):115412, 2012.

\bibitem{calado2015ballistic}
Victor~E Calado, Srijit Goswami, Gaurav Nanda, Mathias Diez, Anton~R Akhmerov,
  Kenji Watanabe, Takashi Taniguchi, Teun~M Klapwijk, and Lieven~MK
  Vandersypen.
\newblock Ballistic josephson junctions in edge-contacted graphene.
\newblock {\em arXiv preprint arXiv:1501.06817}, 2015.

\bibitem{tonnoir2013induced}
Charl{\`e}ne Tonnoir, Amina Kimouche, Johann Coraux, Laurence Magaud, Benjamin
  Delsol, Bruno Gilles, and Claude Chapelier.
\newblock Induced superconductivity in graphene grown on rhenium.
\newblock {\em Physical review letters}, 111(24):246805, 2013.

\bibitem{han2014collapse}
Zheng Han, Adrien Allain, Hadi Arjmandi-Tash, Konstantin Tikhonov, Mikhail
  Feigel’Man, Benjamin Sac{\'e}p{\'e}, and Vincent Bouchiat.
\newblock Collapse of superconductivity in a hybrid tin-graphene josephson
  junction array.
\newblock {\em Nature Physics}, 10(5):380--386, 2014.

\bibitem{ugeda2015}
Miguel~M Ugeda, Aaron~J Bradley, Yi~Zhang, Seita Onishi, Yi~Chen, Wei Ruan,
  Claudia Ojeda-Aristizabal, Hyejin Ryu, Mark~T Edmonds, Hsin-Zon Tsai, et~al.
\newblock Characterization of collective ground states in single-layer nbse2.
\newblock {\em Nature Physics}, 12(1):92--97, 2016.

\bibitem{cao2015quality}
Y.~Cao, A.~Mishchenko, G.~L. Yu, E.~Khestanova, A.~P. Rooney, E.~Prestat, A.~V.
  Kretinin, P.~Blake, M.~B. Shalom, C.~Woods, J.~Chapman, G.~Balakrishnan,
  I.~V. Grigorieva, K.~S. Novoselov, B.~A. Piot, M.~Potemski, K.~Watanabe,
  T.~Taniguchi, S.~J. Haigh, A.~K. Geim, and R.~V. Gorbachev.
\newblock Quality heterostructures from two-dimensional crystals unstable in
  air by their assembly in inert atmosphere.
\newblock {\em Nano Letters}, 15(8):4914--4921, 2015.

\bibitem{efetov2015specular}
DK~Efetov, L~Wang, C~Handschin, KB~Efetov, J~Shuang, R~Cava, T~Taniguchi,
  K~Watanabe, J~Hone, CR~Dean, et~al.
\newblock Specular interband andreev reflections in graphene.
\newblock {\em arXiv preprint arXiv:1505.04812}, 2015.

\bibitem{yang2014superconducting}
S-L Yang, JA~Sobota, CA~Howard, CJ~Pickard, Makoto Hashimoto, DH~Lu, S-K Mo,
  PS~Kirchmann, and Z-X Shen.
\newblock Superconducting graphene sheets in cac6 enabled by phonon-mediated
  interband interactions.
\newblock {\em Nature communications}, 5, 2014.

\bibitem{chapman2015superconductivity}
J~Chapman, Y~Su, CA~Howard, D~Kundys, A~Grigorenko, F~Guinea, AK~Geim,
  IV~Grigorieva, and RR~Nair.
\newblock Superconductivity in ca-doped graphene.
\newblock {\em Nature Physics}, 5, 2014.

\bibitem{rahnejat2011charge}
KC~Rahnejat, CA~Howard, NE~Shuttleworth, SR~Schofield, K~Iwaya,
  CF~Hirjibehedin, Ch~Renner, G~Aeppli, and M~Ellerby.
\newblock Charge density waves in the graphene sheets of the superconductor
  $cac_6$.
\newblock {\em Nature communications}, 2:558, 2011.

\bibitem{yazyev2007defect}
Oleg~V Yazyev and Lothar Helm.
\newblock Defect-induced magnetism in graphene.
\newblock {\em Physical Review B}, 75(12):125408, 2007.

\bibitem{Palacios08}
J.~J. Palacios, J.~Fern\'andez-Rossier, and L.~Brey.
\newblock Vacancy-induced magnetism in graphene and graphene ribbons.
\newblock {\em Phys. Rev. B}, 77:195428, May 2008.

\bibitem{nair12}
RR~Nair, M~Sepioni, I-Ling Tsai, O~Lehtinen, J~Keinonen, AV~Krasheninnikov,
  T~Thomson, AK~Geim, and IV~Grigorieva.
\newblock Spin-half paramagnetism in graphene induced by point defects.
\newblock {\em Nature Physics}, 8(3):199--202, 2012.

\bibitem{santos2012universal}
Elton~JG Santos, Andr{\'e}s Ayuela, and Daniel S{\'a}nchez-Portal.
\newblock Universal magnetic properties of sp3-type defects in covalently
  functionalized graphene.
\newblock {\em New Journal of Physics}, 14(4):043022, 2012.

\bibitem{boukhvalov2008hydrogen}
DW~Boukhvalov, MI~Katsnelson, and AI~Lichtenstein.
\newblock Hydrogen on graphene: Electronic structure, total energy, structural
  distortions and magnetism from first-principles calculations.
\newblock {\em Physical Review B}, 77(3):035427, 2008.

\bibitem{soriano2010hydrogenated}
D~Soriano, Federico Mu{\~n}oz-Rojas, J.~Fern{\'a}ndez-Rossier, and JJ~Palacios.
\newblock Hydrogenated graphene nanoribbons for spintronics.
\newblock {\em Physical Review B}, 81(16):165409, 2010.

\bibitem{HIvan}
Ivan Brihuega.
\newblock Atomic scale control of graphene magnetism using hydrogen atoms.
\newblock {\em SPICE workshop, Magnetic adatoms as building blocks for quantum
  magnetism}, 2015.

\bibitem{nair2013dual}
RR~Nair, I-L Tsai, M~Sepioni, O~Lehtinen, J~Keinonen, AV~Krasheninnikov,
  AH~Castro Neto, MI~Katsnelson, AK~Geim, and IV~Grigorieva.
\newblock Dual origin of defect magnetism in graphene and its reversible
  switching by molecular doping.
\newblock {\em Nature communications}, 4, 2013.

\bibitem{yndurain2014effect}
Felix Yndurain.
\newblock Effect of hole doping on the magnetism of point defects in graphene:
  A theoretical study.
\newblock {\em Physical Review B}, 90(24):245420, 2014.

\bibitem{wehling2007local}
TO~Wehling, AV~Balatsky, MI~Katsnelson, AI~Lichtenstein, K~Scharnberg, and
  R~Wiesendanger.
\newblock Local electronic signatures of impurity states in graphene.
\newblock {\em Physical Review B}, 75(12):125425, 2007.

\bibitem{beenakker2006specular}
CWJ Beenakker.
\newblock Specular andreev reflection in graphene.
\newblock {\em Physical review letters}, 97(6):067007, 2006.

\bibitem{jacob2010orbital}
D~Jacob and G~Kotliar.
\newblock Orbital selective and tunable kondo effect of magnetic adatoms on
  graphene: Correlated electronic structure calculations.
\newblock {\em Physical Review B}, 82(8):085423, 2010.

\bibitem{peres2006electronic}
NMR Peres, F~Guinea, and AH~Castro Neto.
\newblock Electronic properties of disordered two-dimensional carbon.
\newblock {\em Physical Review B}, 73(12):125411, 2006.

\bibitem{wehling2008local}
TO~Wehling, HP~Dahal, AI~Lichtenstein, and AV~Balatsky.
\newblock Local impurity effects in superconducting graphene.
\newblock {\em Physical Review B}, 78(3):035414, 2008.

\bibitem{mcmillan1968tunneling}
WL~McMillan.
\newblock Tunneling model of the superconducting proximity effect.
\newblock {\em Physical Review}, 175(2):537, 1968.

\bibitem{chrestin1997evidence}
A~Chrestin, T~Matsuyama, and U~Merkt.
\newblock Evidence for a proximity-induced energy gap in nb/inas/nb junctions.
\newblock {\em Physical Review B}, 55(13):8457, 1997.

\bibitem{san2015majorana}
P.~San-Jose, J.~L. Lado, R.~Aguado, F.~Guinea, and J.~Fern\'andez-Rossier.
\newblock Majorana zero modes in graphene.
\newblock {\em Phys. Rev. X}, 5:041042, Dec 2015.

\bibitem{lim2015shiba}
Jong~Soo Lim, Rosa L{\'o}pez, Ram{\'o}n Aguado, et~al.
\newblock Shiba states and zero-bias anomalies in the hybrid
  normal-superconductor anderson model.
\newblock {\em Physical Review B}, 91(4):045441, 2015.

\bibitem{matsuura1977effects}
Tamifusa Matsuura.
\newblock The effects of impurities on superconductors with kondo effect.
\newblock {\em Progress of Theoretical Physics}, 57(6):1823--1835, 1977.

\bibitem{nagaoka1971bound}
Yosuke Nagaoka and Tamifusa Matsuura.
\newblock Bound states due to a magnetic impurity in a superconductor. i.
\newblock {\em Progress of Theoretical Physics}, 46(2):364--386, 1971.

\bibitem{soda1967sd}
Toshio Soda, Tamifusa Matsuura, and Yosuke Nagaoka.
\newblock sd exchange interaction in a superconductor.
\newblock {\em Progress of Theoretical Physics}, 38(3):551--567, 1967.

\bibitem{satori1992numerical}
Koji Satori, Hiroyuki Shiba, Osamu Sakai, and Yukihiro Shimizu.
\newblock Numerical renormalization group study of magnetic impurities in
  superconductors.
\newblock {\em Journal of the Physical Society of Japan}, 61(9):3239--3254,
  1992.

\bibitem{bauer2007spectral}
Johannes Bauer, A~Oguri, and AC~Hewson.
\newblock Spectral properties of locally correlated electrons in a
  bardeen--cooper--schrieffer superconductor.
\newblock {\em Journal of Physics: Condensed Matter}, 19(48):486211, 2007.

\bibitem{kirvsanskas2015yu}
Gediminas Kir{\v{s}}anskas, Moshe Goldstein, Karsten Flensberg, Leonid~I
  Glazman, and Jens Paaske.
\newblock Yu-shiba-rusinov states in phase-biased superconductor--quantum
  dot--superconductor junctions.
\newblock {\em Physical Review B}, 92(23):235422, 2015.

\bibitem{haase2011magnetic}
P~Haase, S~Fuchs, T~Pruschke, H~Ochoa, and F~Guinea.
\newblock Magnetic moments and kondo effect near vacancies and resonant
  scatterers in graphene.
\newblock {\em Physical Review B}, 83(24):241408, 2011.

\bibitem{sofo2012magnetic}
JO~Sofo, Gonzalo Usaj, PS~Cornaglia, AM~Suarez, AD~Hern{\'a}ndez-Nieves, and
  CA~Balseiro.
\newblock Magnetic structure of hydrogen-induced defects on graphene.
\newblock {\em Physical Review B}, 85(11):115405, 2012.

\end{thebibliography}

\end{document}